\documentclass[lettersize,journal]{IEEEtran}
\usepackage{amsmath,amsfonts}
\usepackage[ruled,vlined]{algorithm2e}
\usepackage{array}
\usepackage{subcaption}

\usepackage{outlines}
\usepackage{caption}
\usepackage{multirow}
\usepackage{hyperref}
\usepackage{comment}
\usepackage{enumerate}
\usepackage{booktabs}

\usepackage{textcomp}
\usepackage{stfloats}
\usepackage{url}
\usepackage{verbatim}
\usepackage{graphicx}
\usepackage{cite}
\begin{document}

\title{Time-Series Analysis Approach for Improving Energy Efficiency of a Fixed-Route Vessel in Short-Sea Shipping}

\author{Mohamed Abuella,~\IEEEmembership{} Hadi Fanaee,~\IEEEmembership{} Slawomir Nowaczyk,~\IEEEmembership{} 
Simon Johansson,~\IEEEmembership{}and Ethan Faghani.~\IEEEmembership{}

\thanks{Manuscript received November 1, 2023; revised November 16, 2023.}

\thanks{This work was supported by the Sweden's innovation agency (Vinnova).(\textit{Corresponding author: Mohamed Abuella}.)\\
Mohamed Abuella, Hadi Fanaee, and Slawomir Nowaczy are with Center for Applied Intelligent Systems Research (CAISR), Halmstad University, Kristian IV:s väg 80523, Halmstad, 30118 Sweden (e-mail: mohamed.abuella@hh.se;
hadi.fanaee@hh.se; Slawomir.Nowaczyk@hh.se).
Simon Johansson and Ethan Faghani are with CetaSol AB, Gothenburg, 41251, Sweden (e-mail: simon.johansson and ethan.faghani\}@cetasol.com)}}

\markboth{Journal of \LaTeX\ Class Files,~Vol.~14, No.~8, August~2021}%
% \markboth{This is a preprint of a potential publishable work for IEEE TRANSACTIONS and JOURNALS.}%
{Shell \MakeLowercase{\textit{et al.}}: A Sample Article Using IEEEtran.cls for IEEE Journals}

% % Remember, if you use this you must call \IEEEpubidadjcol in the second column for its text to clear the IEEEpubid mark.

\maketitle
\begin{abstract}
Several approaches have been developed for improving the ship energy efficiency, thereby reducing operating costs and ensuring compliance with climate change mitigation regulations. Many of these approaches will heavily depend on measured data from onboard IoT devices, including operational and environmental information, as well as external data sources for additional navigational data.
In this paper, we develop a framework that implements time-series analysis techniques to optimize the vessel's speed profile for improving the vessel's energy efficiency.
We present a case study involving a real-world data from a passenger vessel that was collected over a span of 15 months in the south of Sweden. 
The results indicate that the implemented models exhibit a range of outcomes and adaptability across different scenarios.
The findings highlight the effectiveness of time-series analysis approach for optimizing vessel voyages within the context of constrained landscapes, as often seen in short-sea shipping.
\end{abstract}


\section{Introduction}\label{Sec1_Intro}
\IEEEPARstart{T}{he} transportation of commercial freight by sea is the most environment-friendly transportation mode, since it emits less greenhouse gases (GHGs) per tonne of capacity and kilometer of distance, thus making resulting a smaller carbon footprint and lower impact on global climate. 
Short-Sea Shipping (SSS) is a commercial transportation mode that does not involve intercontinental cross-ocean. The SSS provides a cost-efficient and environment-friendly alternative for transportation by utilizing inland and coastal waterways to transport the commercial freight \cite{sss_def}. 

On the other hand, the SSS produces some negative effects on the natural habitats and polluting the air along the coasts of populated cites~\cite{donner2018}. As a response to this, the International Maritime Organization (IMO) have conducted many studies and recommended standards and imposed policies for the maritime sector to reduce the carbon dioxide (CO$_2$) to 40\% by 2030 and cut 50\% of all GHGs by 2050, based on the emissions in 2008~\cite{Ampah2021}.

Furthermore, COVID-19 pandemic has accelerated the digitalization of the entire shipping industry globally, and hence attracted a  profound consideration to data collection and preparation stages~\cite{world2020accelerating}.
The operational and some environmental conditions can be accessed through an Automatic Identification System (AIS) messages, which is a service developed by the International Maritime Organization (IMO) in 2002 to record the sensor measurements and send the vessel position information for the traffic between other ships and neighboring shores~\cite{C2020}.

In a broader perspective, for improving the vessel's energy efficiency and harnessing more fuel savings and less GHG emissions, there are mainly two procedures. The first strategy is in the ship design stage, where the ship is built to obtain a body and equipped machinery that work efficiently. The second strategy is during the ship operation over the water or at the ports. This latter procedure can be achieved by adopting energy management plans that optimally enhance the energy efficiency and fuel consumption~\cite{zis2020ship}.

This paper proposes a data-driven framework for optimizing vessel speed profiles to improve energy efficiency in short-sea shipping. \\
The main contributions of this paper can be summarized as follow:
\begin{itemize}
    \item Modeling of energy efficiency: Develop a data-driven model for voyage energy efficiency, including:\\
    A spatiotemporal aggregation of operation and navigation data from onboard and external sources to capture the impact of both spatial and temporal factors on voyage energy efficiency. \\
    Introduce an efficiency score that considers both total fuel consumption and voyage duration to measure the voyage energy efficiency.
    \item Data clustering: Clustering the data of voyages and sorting them based on their efficiency scores. This clustering enables the voyage optimization algorithm to learn more insights for better actions, either by selecting the best voyages or by eliminating the worst voyages.
    \item {Time-series analysis models and comparative analysis: Four time-series based models are implemented as algorithms of voyage speed optimization. Then, a rigorous evaluation of their performance is conducted across different data clusters and using metrics that account for voyage efficiency.
    \item Practical implication: Demonstrate the significant effectiveness and practicality of the proposed approach for fixed-route vessels in short sea shipping, where the options for obtaining efficient voyages are limited. The approach also aligns with the guidance of domain experts, adhering to safety and traffic considerations.}
\end{itemize}
The remainder of the paper is organized as follows: related work is presented in Section~\ref{Sec_relatedwork}. The methodology of the proposed framework is covered in Section~\ref{Sec_method}. Section~\ref{Sec_case_study} introduces the case study, including a description of the data used and the framework's implementation. The results and outcomes of the models are discussed in Section~\ref{Sec_results}. Finally, conclusions are addressed in Section~\ref{Sec_conclusion}.

\section{Related Work}\label{Sec_relatedwork}

Various techniques can be applied for the voyage optimization problem, as demonstrated in prior studies such as~\cite{chen2021art, Walther2016, fan2022review, moradi2022}. Some of these approaches have been implemented specifically for short-sea shipping, as indicated in the domain of~\cite{zakaria2022instruments} and~\cite{grifoll2018potential}.

Different modeling and optimization algorithms for ship weather routing have been investigated thoroughly in~\cite{walther2016modeling}. It was found that the effectiveness of these algorithms strongly depends on specific requirements concerning the objectives, control variables and constraints as well as the implementation.\\
The majority of these voyage optimization approaches are applied to ocean-crossing ships by controlling mainly the vessel speed and its route. Whereas, in the SSS, especially for fixed-route vessels like ours, there are fewer options available for voyage optimization.
Therefore, the scope of this paper is mainly focused on short-sea shipping and the pertinent research literature.

In the third IMO GHG study 2014~\cite{IMO2014third}, it was assumed that weather effects alone would be responsible for 15\% of additional power margin on top of the theoretical propulsion requirements of ocean-going ships, and a 10\% additional power requirement for coastal ships. \\
In a recent adaption of the Ship Traffic Emission Assessment Model (STEAM), propelling power is determined by wave height and directions, accounting for the environmental conditions in a highly detailed manner~\cite{C2020}.

Recent research studies~\cite{Bellingmo2021, jorgensen2022ship} have explored energy-efficient routing for an electric ferry in Western Norway. They rely on operational data from onboard measurements and environmental conditions from the Norwegian Meteorological Institute, and proposed a hybrid physics-guided machine learning model for optimizing the ship route. Based on their findings, the hybrid model was showing an energy reduction of 3.7\% compared to the actual consumption, simply by applying minor route and speed profile alterations as guided by the provided weather forecasts.

Researchers from Napa Ltd. in Finland conducted several studies on voyage optimization. In two of their studies~\cite{sugimotodigital} and~\cite{haranen2017role}.
These experts stated that their products of voyage optimization can achieve a fuel cost reduction of more than 10\% with 2\% to 4\% savings from trim optimization and 6\% to 8\% from speed and route optimization.~\cite{Wingrove2016}.

In the study conducted by Huotari et al.~\cite{huotari2021convex}, where they used a combined model with both dynamic programming and convex optimisation to obtain optimal speed profiles. The fuel savings were around 1.1\% and for voyages with substantial variance in environmental conditions, the fuel savings reached as high as 3.5\%.

In the review paper by Wang et al.~\cite{wang2022comprehensive} numerous studies are explored, including coastal and inland shipping, also revealing a diverse range of fuel savings outcomes.\\
Meanwhile, regarding weather routing, specifically through speed and route optimization, it has been demonstrated in~\cite{walther2021development} that the potential savings of carbon dioxide emissions and fuel costs are in range of 5\% to 10\%.

%%%%%%%%%%%%
In accordance with the main contributions highlighted in Section (\ref{Sec1_Intro}), our paper distinguishes itself from previous research by tackling the distinctive operational challenges of SSS, employing time-series analysis models for optimizing voyage speed, and conducting a comprehensive comparative analysis to rigorously evaluate the robust performance of these models across diverse data clusters.
  
%%%%%%%%%%%%%%%%%%%%%%%%%%%%%%%%%%%%%%%%%%%
\section{Methodology}\label{Sec_method}

The main objective of this study is to improve the vessel voyage by optimizing its speed to enhance the vessel's energy efficiency. In other words, reducing the vessel's fuel consumption within constrained arrival time.

\subsection{Framework of Voyage Optimization}\label{sec_framework}
\begin{figure}[!ht]
\centering
\includegraphics[width=0.5\linewidth]{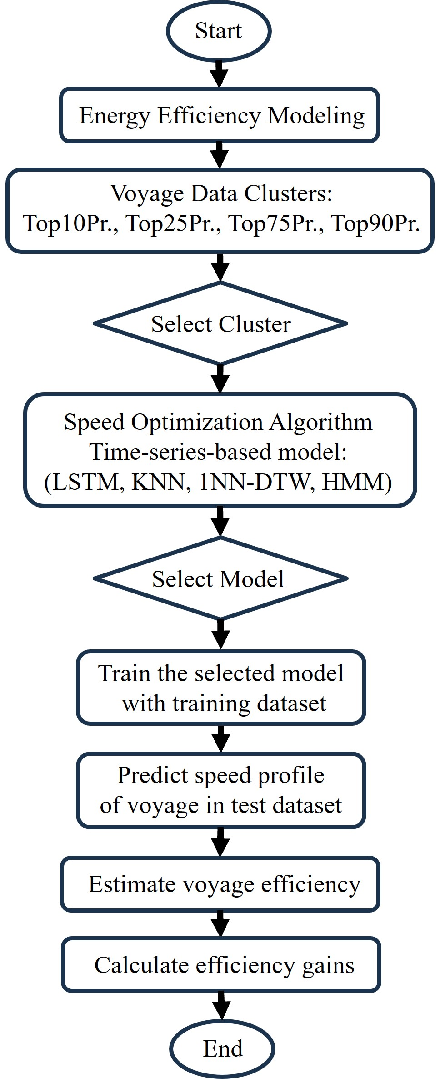}
\caption{Framework of vessel voyage optimization.}
\label{fig_Fuel_Min_problem}
\end{figure}

The framework of the developed approach for improving the vessel voyages is depicted in Figure~\ref{fig_Fuel_Min_problem}.

%%%%%%%%%%%%%%%
Data Processing and Clustering is a vital component of our framework. 
The primary objective is to identify and sort the voyages based on their efficiency scores.
Then, train the models with the sorted data clusters iteratively, to distill insights from the voyages with different behaviours. 
Thus, the trained models will gain valuable insights into the performance and operational patterns of the vessel.
%%%%%%%%%%%%%%%
\subsection{Energy Efficiency Modeling}\label{sec_EE_Modeling}

This part presents a mathematical and visual overview of the fundamental theoretical background that forms the basis for modeling vessel energy efficiency---an indispensable element within our comprehensive framework. \\

To estimate the vessel's energy efficiency in this framework of voyage optimization, we employed a previously developed model equipped with artificial intelligence (XAI) and machine learning techniques. More details about this energy efficiency modeling approach can be found in~\cite{abuella2023xai}.

The efficiency score ($\mathrm{Eff_{score}}$) is calculated from the normalized total fuel and time for every voyage, as following:
\begin{equation}\label{eq_effsocre}
\mathrm{Eff_{Score}} = 1- \frac{2 \times  [Fuel_{Tl_{Nm}} \times Time_{Tl_{Nm}}]} {[Fuel_{Tl_{Nm}} + Time_{Tl_{Nm}}]}
\end{equation}

The efficiency score considers the proportional reduction in both fuel consumption and time, assessing the vessel's efficient use of resources during the voyage.
%%%%%%%%%%%%%%%%%%%%%%%%%%%%%%%%%%%%%%%%%%%%%%%%%%%%%%

\begin{figure}[!ht]
\centering
\begin{subfigure}{0.34\linewidth}
    \includegraphics[width=\linewidth]{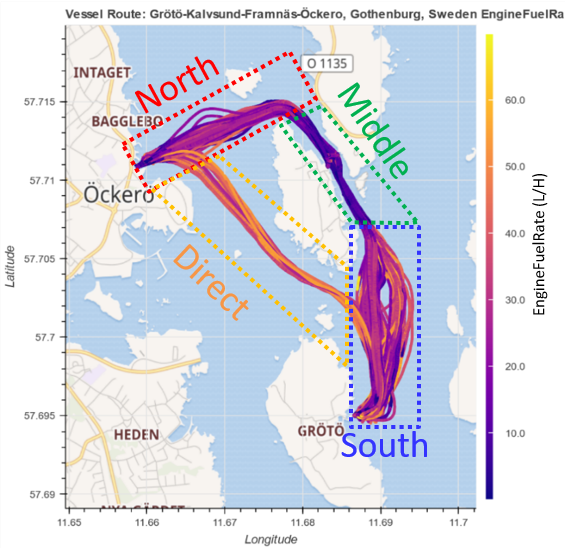}
    \caption{Vessel's route.}
    \label{Aggreg_rts}
\end{subfigure}
\hfill
\centering
\begin{subfigure}{0.64\linewidth}
\centering
   \includegraphics[width=\linewidth]{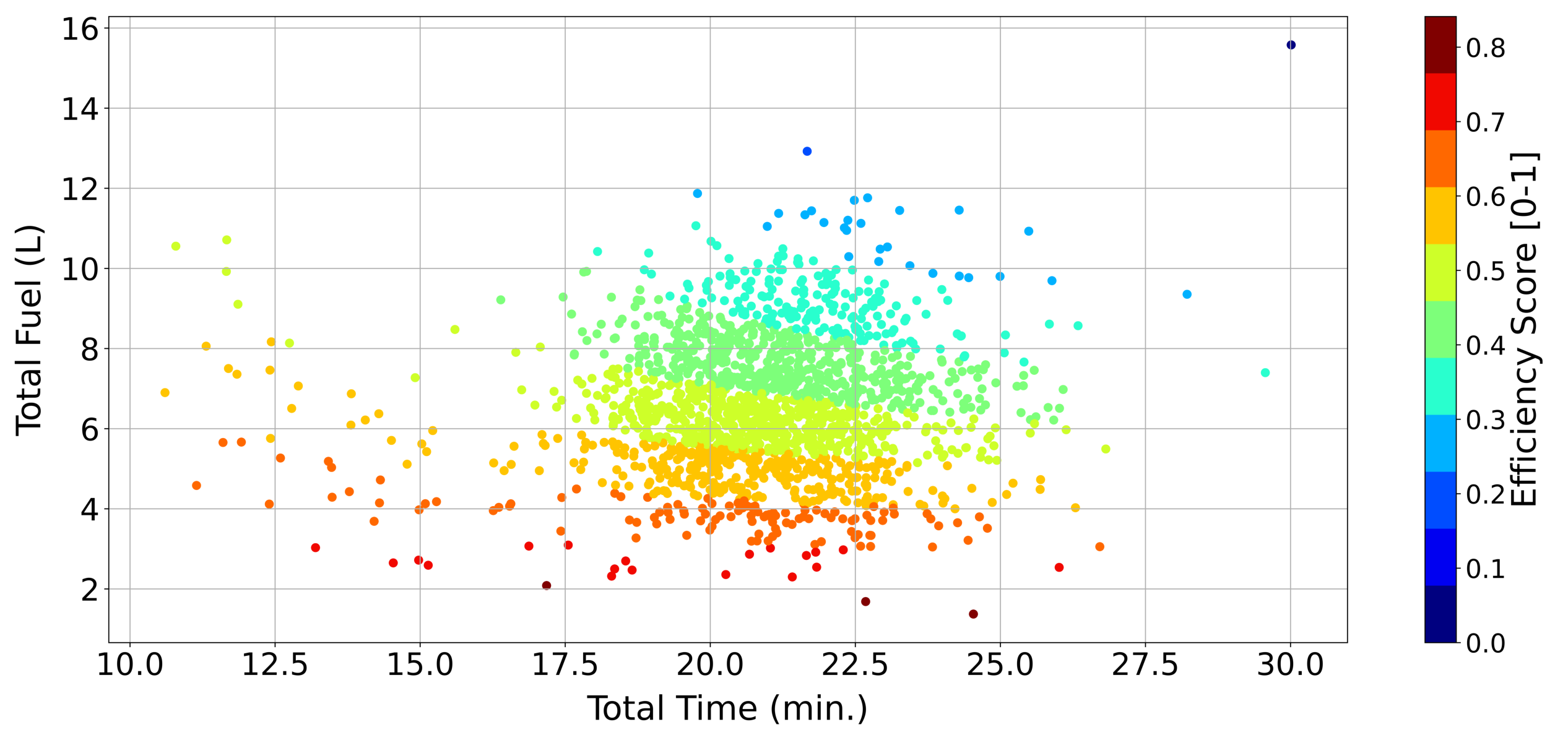}
\caption{Data are projected by Efficiency Score vs. fuel and time.}
\label{Global_Eff_vs_FL_Time}
\end{subfigure}
\caption{The vessel voyages and the aggregated data projected by the Efficiency Score.}
\label{Aggregation_Data}
\end{figure}

Figure~\ref{Aggregation_Data} facilitates to visualize the process of aggregation for the vessel's voyages. First by illustrating the voyages in terms of space (i.e., latitude and longitude) as shown in Figure~\ref{Aggreg_rts}, and second by representing the aggregated voyages as points in new dimensions of Efficiency Score versus total fuel and total time. The aggregated data and its new dimensions are projected as in Figure~\ref{Global_Eff_vs_FL_Time}.

The representation of voyages in terms of fuel and time is done by adopting the concept of the Efficiency Score (Eff-Score). The efficiency scores for all vessel's voyages are presented in Figure~\ref{Global_Eff_vs_FL_Time}.
It is evident that the voyage with lower fuel and shorter time have higher efficiency scores, and vice versa.

\begin{figure}[!ht]
\centering
\centering
\includegraphics[width=\linewidth]{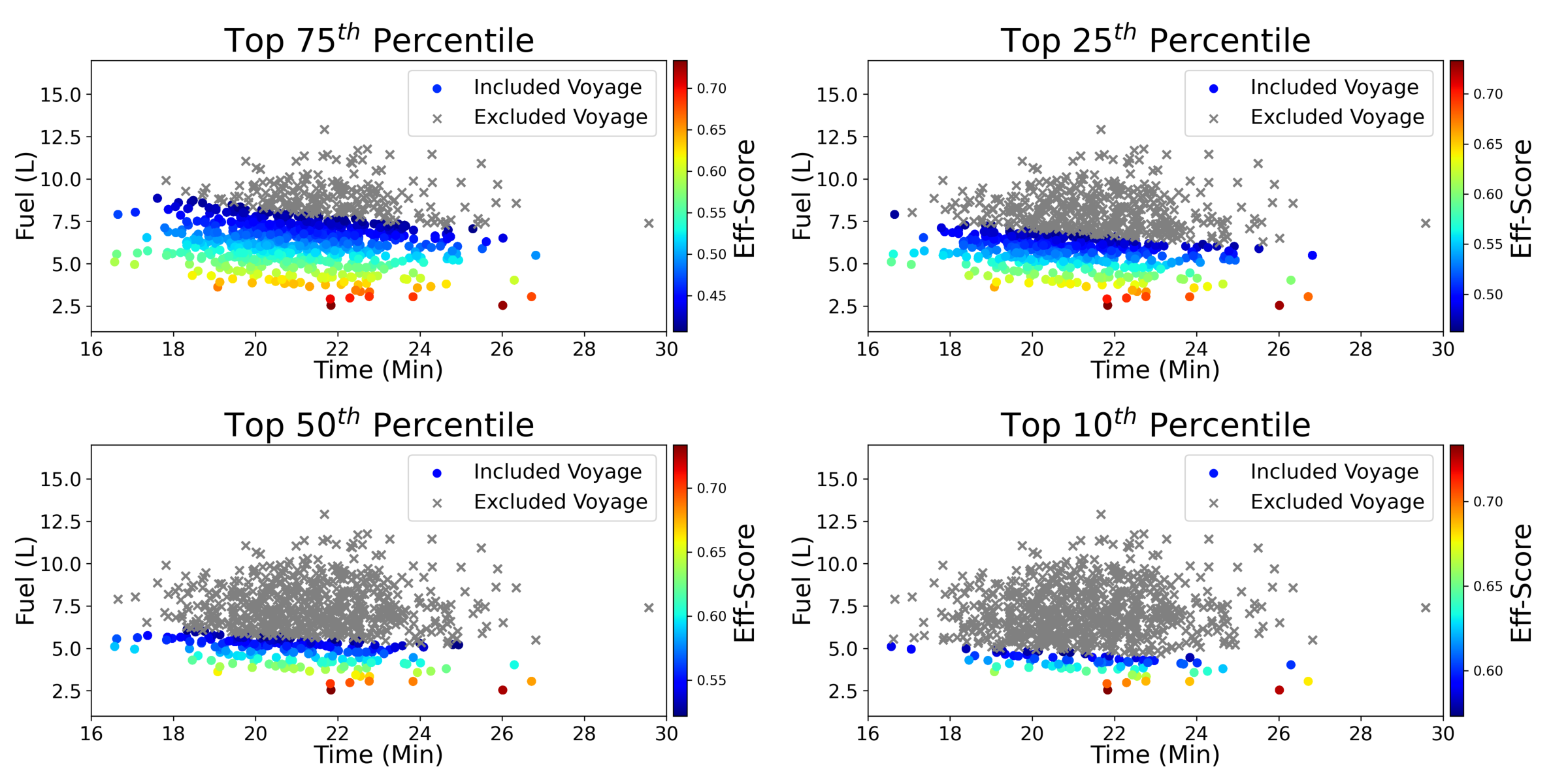}
\caption{Four clusters of voyages based on their efficiency.}
\label{vog_clusters}
\end{figure}

%%%%%%%%%%%%%%%%%
\begin{algorithm}
\caption{Modeling of Energy Efficiency and Clustering of Voyages Data Based on Their Energy Efficiency}
\label{data_proc_Alg}

\KwData{Voyages data of the vessel}
\KwResult{Clusters of Voyages}
{Load the operational and navigational data, including speed, course, fuel, position, distance, and weather\;
Tag the datapoints to its corresponding voyage, $V_{id}$\;}
\ForEach{voyage $V_{i}$ in voyages data}{
    Calculate total fuel consumption and time for $V_{i}$\;
    Normalize total fuel and time for $V_{i}$ based on their maximum values of all voyages\;
    Calculate the \textit{Eff-Score} as described in Eq.~\eqref{eq_effsocre}, and assign it to all datapoints of this voyage $V_{i}$\;}

Initialize four empty lists for each cluster: $Top75Pr$, $Top50Pr$, $Top25Pr$, $Top10Pr$ (Percentiles of Eff-Scores)\;

\ForEach{data point in all data}{
    Extract the Eff-Score of the data point\;
    \If{Eff-Score is in the top 75\%}{
        Append the data point to $Top75Pr$\;
    }
    \ElseIf{Eff-Score is in the top 50\%}{
        Append the data point to $Top50Pr$\;
    }
    \ElseIf{Eff-Score is in the top 25\%}{
        Append the data point to $Top25Pr$\;
    }
    \ElseIf{Eff-Score is in the top 10\%}{
        Append the data point to $Top10Pr$\;
    }
}
\end{algorithm}
%%%%%%%%%%%%%%%%%
There are four voyage data clusters, namely $Top75Pr$, $Top50Pr$, $Top25Pr$, and $Top10Pr$, as shown in Figure~\ref{vog_clusters}, These clusters are categorized on their respective Eff-Score percentiles, enabling a structured analysis of voyage efficiency across various percentile groups.

%%%%%%%%%%%%%%%%%%%%%%%%%%%%%%%%%%%%%%%%%%%%%%%%%%%%%%%%%%%%
\subsection{Time-Series Analysis Models for Voyage Optimization}\label{sec_TSA_models}
For the purpose of vessel voyage optimization, we need a model to optimize the vessel's speed profile for improving the vessel's energy efficiency.
This ideal model should mainly be able to:
\begin{itemize}
    \item Model the temporal dependencies in vessel speed profiles.
    \item Incorporate external variables, such as weather conditions, into the modeling process.
    \item Adapt to changing conditions to provide real-time optimized speed profiles.
\end{itemize}

In order to meet these requirements, we adopt several time-series analysis models, which are described with some details as follow.

\subsubsection{Long Short-Term Memory (LSTM)}\label{Sec_lstm}
It is a type of recurrent neural network (RNN) architecture designed for sequential data analysis. It is particularly useful for modeling and predicting time-series data, making it a valuable tool for improving energy efficiency in vessel operations~\cite{hochreiter1997long}.\\
LSTM networks are well-suited for tasks involving time-series sequences of varying lengths, capturing long-term dependencies, and handling irregular temporal patterns. The LSTM architecture deals specifically to address the problem of vanishing gradients that often occurs in other RNN structures~\cite{staudemeyer2019understanding}.

%%%%% this could be replacced in the resutls and discussion section  %%%%%
The LSTM's capability to learn the hidden insights from selected data, particularly in terms of vessel's efficiency scores, comes with a trade-off. The data clustering may result in a smaller dataset, which potentially limiting the LSTM's deep learning capabilities. Additionally, data clustering increases the susceptibility of LSTM to overfitting.\\
LSTM can be employed to model and predict the vessel's speed to improve its fuel efficiency, taking into account various factors such as weather conditions and operational parameters 
%%%%%%%%%%%%%
\subsubsection{k-Nearest Neighbors (kNN)}\label{Sec_knn}
The kNN algorithm is a non-parametric model that makes no assumptions and operates on the principle of determining an unknown observation's class by measuring distances to nearby observations, attributing the observation to the majority class of its nearest neighbors. The parameter k represents the number of neighboring observations considered when classifying a given observation, and its value is determined through a search for the optimal choice that maximizes accuracy in the training set.

The paper by Cover and Hart~\cite{cover1967nearest} is a foundational work in pattern recognition and machine learning, introduced the KNN algorithm for pattern classification by utilizing the nearest neighbors to classify data points based on majority vote.

KNN can be employed to analyze and predict energy consumption patterns based on similar historical data. By considering the nearest neighbors of a given operational scenario, KNN provides a straightforward approach to making energy-efficient decisions. The choice of the parameter 'k,' representing the number of neighbors, plays a crucial role in the accuracy of KNN-based models.
%%%%%%%%%%%%%%%
\subsubsection{1-Nearest Neighbor with Dynamic Time Warping (1NN-DTW) }\label{Sec_dtw}
This model is a distinctive variant of kNN with (k=1) and with a distance-wise measure that utilizes Dynamic Time Warping (DTW) algorithm for a comparison of two time-series sequences that may have different lengths, time shifts, and speed variations~\cite{berndt1994using, keogh2005exact, ding2008querying, rakthanmanon2012searching, bagnall2017great}.
The 1NN-DTW algorithm can be utilized to measure the similarity between the given speed profile and other measured vessel speed profiles with higher efficiency scores. 
%DTW can also incorporate external factors (i.e, exogenous variables), such as the environmental conditions, into the similarity calculation.
1NN-DTW is capable to capture the temporal dependency of speed profiles, even in short-sea shipping where the options of control the ship to improve its energy efficiency are limited.

Thus, the 1NN-DTW identifies the efficient observed speed profile that most similar to the given speed profile to be chosen as suggested an optimized speed profile for the current journey of the vessel that can lead to improved energy efficiency.
%%%%%%%%%%%%%%%
\subsubsection{Hidden Markov Model (HMM)}\label{Sec_hmm}
This model is based on probabilistic modeling and use a variety of techniques from the statistical modeling, and they are widely used in time-series analysis and pattern recognition.\\
The work by Rabiner in 1989~\cite{rabiner1989tutorial} is an essential reference in the application of HMM. This tutorial delved into the formulation of a statistical method for representing speech, showcasing a successful HMM system implementation with a focus on discrete or continuous density parameter distributions.

In the context of improving energy efficiency in constrained environments, HMMs can be applied to model the underlying patterns and transitions in ship operational data, allowing for more informed decisions on optimizing energy consumption.\\
The HMM estimates three main weather states and dynamically adjusts the SpeedOverGround (SOG) predictions. During calm weather, it selects the maximum observed speed in the historical data at that condition. In moderate conditions, it uses the average speed, while in rough weather, it relies on the minimum speed profiles. 
For instance, Eq.~\ref{Eq_sog_rough} indicates that the predicted speed (SOG$_{\text{Pred}_i}$) by HMM is the minimum value of measured speed profiles in rough weather state, and likewise for other weather states.

\begin{equation}\label{Eq_sog_rough}
SOG_{\text{Pred.}_i} = \min(SOG_{\text{Meas.}} \, | \, \text{Rough weather})
\end{equation}
This adaptability to changing weather conditions enhances voyage optimization, making it a crucial component of the overall framework.\\
%%%%%%%%%%%%%%%%%%%%%%%%%%%%%%%%%%%%%%%%%%%%%%%%%
\section{Case Study}\label{Sec_case_study}

\subsection{Data Collection}\label{sec_data_col}
The ship’s onboard data have been received from our industry partner CetaSol AB in Gothenburg~\cite{CetaSol2022}.  The data has been gathered over a period of 15 months, between January 2020 and March 2021. It has a 3Hz frequency and records about the ship’s position, course direction, and speed. It is also including some of operational and meteorological data, such as fuel rate, engine speed, torque, acceleration, wind speed and direction.

Some information about the ship and its voyage can be found on Marine Traffic website~\cite{marinetrafffic}.\\
Other weather variables such as wave height and sea current speed and direction have been collected from external sources, Copernicus Marine Service~\cite{Copernicus2020} and Stormglass~\cite{StormGlass2022} APIs.

\begin{table}[ht]
\begin{center}
\caption{The navigational variables and their data sources.} \label{vars_abbrev_tab}
\begin{tabular}{cc||cc} 
\hline
Variable & Source & Variable & Source\\
\hline 
Latitude & Onboard & WindSpeed\_cps & Copernicus \\
Longitude & Onboard & WindDirection\_cps & Copernicus \\
SpeedOverGround & Onboard & WaveHeight & Copernicus \\
HeadingMagnetic & Onboard & WaveDirection & Copernicus \\
Pitch & Onboard & WindSpeed\_sg & Stormglass \\
Roll  & Onboard & WindDirection\_sg & Stormglass\\
WindSpeed\_onb & Onboard & CurrentSpeed & Stormglass \\
WindDirection\_onb & Onboard & CurrentDirection & Stormglass\\
\hline
\end{tabular}
\end{center}
\end{table}

\subsection{Data Preparation and Validation}\label{sec_data_valid}
The external weather data are past forecasts (hindcasts), which have reanalysed to become hourly in temporal resolution and with 0.25 to 0.5 degree as a spatial resolution.
Trilinear interpolation in time and space dimensions has been applied on external weather data to be more suitable for time and position frames of the given vessel routing. 
Therefore, the weather and onboard data are used in this analysis with a temporal resolution of 1-minute in average.

\begin{figure}[!ht]
\centering
\centering
\includegraphics[width=\linewidth]{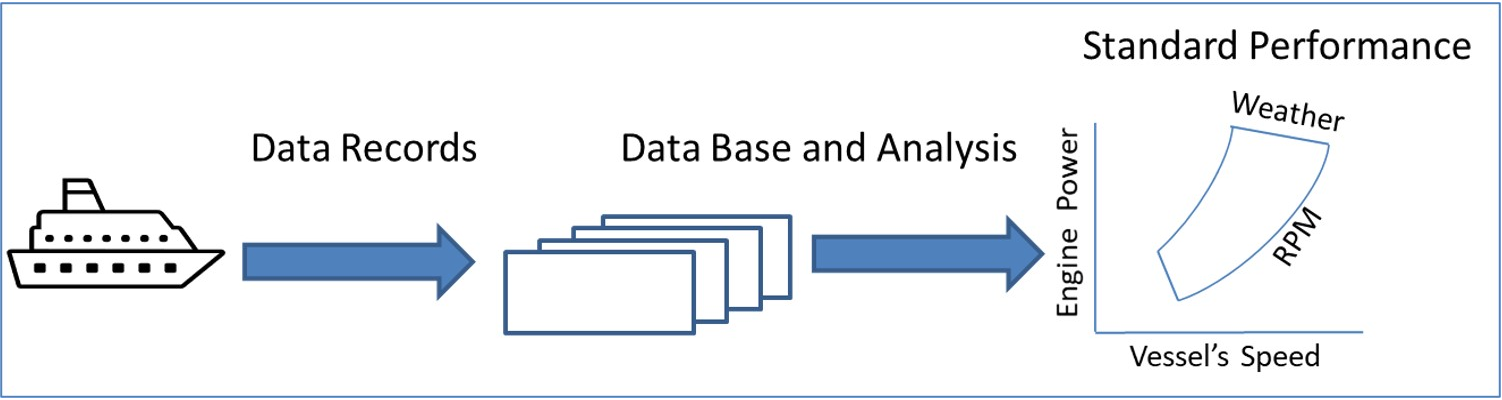}
\caption{Vessel's data analytics and standard performance graph for the case study~\cite{Carlton2018}}
\label{graph_std}
\end{figure}

The data validation is conducted through the cruising-speeds mode is to reduce the other vessel effects on the fuel consumption, and thus, producing graphs that can be then compared with the general ship's standard performance.
The operational and weather data validation is carried out visually, as shown in Figure~\ref{graph_std}.

In a prior publication~\cite{abuella2023xai}, we have addressed the challenges of data representation and energy modeling in short-sea shipping (SSS), by proposing a data-driven modeling approach of voyage efficiency, which combines and aggregates data from multiple sources and seamlessly integrates explainable artificial intelligence (XAI) to attain clear insights about the energy efficiency for a vessel in SSS.

%%%%%%%%%%%%%%%%%
\subsection{Implementation of the Framework}\label{sec:implemen}
The implementation of our approach of a time-series analysis-based voyage optimization framework for a fixed-route vessel of our case study is depicted in Figure~\ref{fig_Fuel_Min_problem}, and the step-by-step process is described by algorithms~\ref{data_proc_Alg} and~\ref{TSA4EE_Alg}.\\
For more detailed information about setting up the models and their specifications including various parameters, you may refer to the source codes, which are developed in Python 3.9.7 to produce the results of this study. These source codes are available at:~\url{https://github.com/MohamedAbuella/TSA4EESSS.}

\begin{algorithm}
\caption{Speed Optimization Models for Improving Voyage Efficiency}
\label{TSA4EE_Alg}

\KwData{Refer to \textbf{Algorithm~\ref{data_proc_Alg}} for data processing and clustering.\\ Add SOG$_{\text{Meas.}}$ and Weathers to Inputs.}

\ForEach{$C_{k}$ in CT (sorted by Eff-Score)}{
    \ForEach{\textbf{model} in [LSTM, KNN, 1NN-DTW, HMM]}{
        Train \textbf{model} with voyages in $C_k$\;
        \ForEach{Voyage $v_{i}$ in test dataset $\notin C_k$ }{
            \If{model is LSTM}{
                Predict SOG$_{\text{Pred.}_i}$ using trained LSTM\;
            }
            \ElseIf{model is KNN}{
                Predict SOG$_{\text{Pred.}_i}$ using trained kNN\;
            }
            \ElseIf{model is 1NN-DTW}{
                Predict SOG$_{\text{Pred.}_i}$ based on the most similar voyage in $C_{k}$\;
            }
            \ElseIf{model is HMM}{
                Predict SOG$_{\text{Pred.}_i}$ based on weather states (Calm, Moderate, Rough)\;
            }
            Estimate efficiency for SOG$_{\text{Meas.}_i}$ and SOG$_{\text{Pred.}_i}$\;
        }
        Evaluate efficiency gains for test voyages\;
    }
}
\end{algorithm}

%%%%%%%%%%%%%%%

%%%%%%%%%%%%%%%%%%%%%%%%%%%%%%%%%%%%%%%%%%%%%%%%%%%%%%

\section{Results and Discussion}\label{Sec_results}
The vessel has a fixed-route which starts from the southern port to the northern port or vice versa.
This route can be divided into four segments, specifically North, Middle, South, and Direct, as depicted in Figure~\ref{Aggreg_rts}.\\
Cruising speeds are more common in North, South, and Direct segments of the vessel's route. Meanwhile, in the Middle segment, the vessel typically operates at maneuvering speeds, due to the presence of two ports located on west and east sides of the canal.

\begin{figure}[!ht]
\centering
\includegraphics[width=\linewidth]{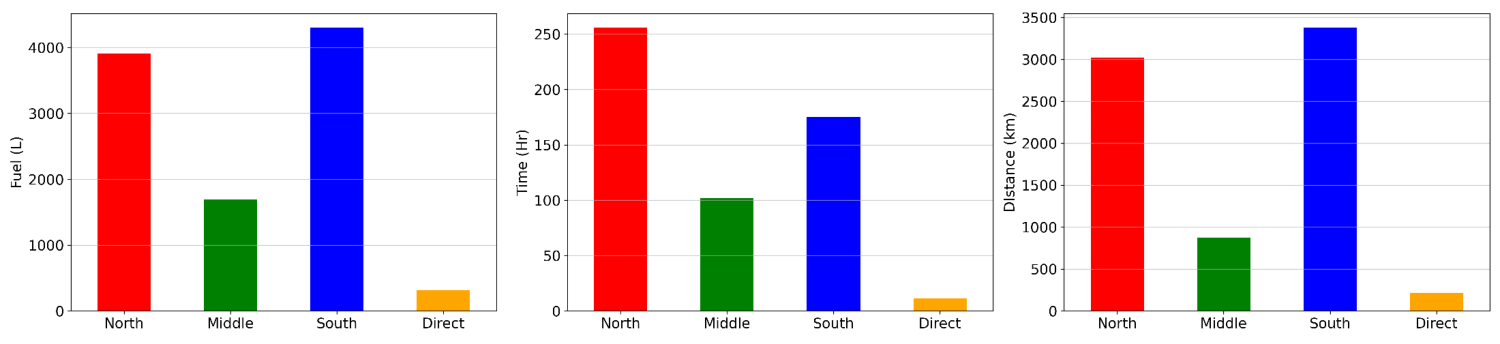}
\caption{Barplots for statistics of Fuel, Time, and Distance in different route segments}
\label{Stats_NMS_sections}
\end{figure}

We have first conducted a statistical analysis on the dataset.
Figure~\ref{Stats_NMS_sections} illustrates some important statistic for all aggregated voyage, with regard to the accumulated fuel, time, and distance at different route segments. The route segments are depicted in Figure~\ref{Aggreg_rts}.

%%%%%%%%%%%%%%%%%%%%%%
As it can be seen from Table~\ref{Stats_Speed_Modes}, the difference of fuel consumption of the and cruising speeds is 5.47\%, so that and also based on the recommendations from domain experts in compliance with maritime regulations including safety and traffic considerations, it might be more practical to primarily focus on optimizing cruising speed.
%%%%%%%%%%%%%%%%%%%
\begin{table}[h]
\centering
\caption{Statistics of the dataset for different speed modes}
\label{Stats_Speed_Modes}
\begin{tabular}{|c|c|c|c|}
\hline
\multirow{2}{*}{Variable} & \multicolumn{2}{c|}{Speeds} & \multirow{2}{*}{Difference (\%)} \\
\cline{2-3}
& All & Cruising & \\
\hline
Fuel, total (Liter) & 1329.2 & 1256.62 & 5.47\% \\
Time, total (Hour) & 48.04 & 22.32 & 53.57\% \\
Distance, total (km) & 608.72 & 349.2 & 42.68\% \\
Speed, average (m/s) & 2.67 & 1.67 & 37.5\% \\
\hline
\end{tabular}
\end{table}
%%%%%%%%%%%%%%%%%%%
After that we have implemented the framework for improving the vessel's energy efficiency, as presented in~\ref{sec:implemen} of Section~\ref{Sec_case_study}.
Then, we evaluate the results, and for a sake of a fair comparison, we are injected both the actual measured and optimized speed profiles into the same estimation model of energy efficiency to predict fuel and time before and after the improving framework of energy efficiency is being implemented. Once the fuel and time estimated, we compute the efficiency to determine how much energy has been saved.

One of our main metrics to evaluate the model performance of voyage efficiency optimization is the gain of efficiency scores, as represented in (\ref{eq:eff_gain}). 
\begin{equation}
Eff._{Gain} = \frac{Eff.Score_{{Pred.}} - Eff.Score_{Meas.}}{Eff.Score_{Meas.}} \times 100
\label{eq:eff_gain}
\end{equation}
Where $Eff.Score_{{Meas.}}$ and $Eff.Score_{{Pred.}}$ represent the voyage efficiency obtained with measured and predicted speed profiles, respectively.

\begin{table}
\centering
\caption{Average efficiency gains (Eff. Gains \%, see Eq.~\ref{eq:eff_gain}) and counts of improved voyages (IV Count\#) out of 162 voyages in the test dataset.}
\label{Table_Eff_gain_couts}
\begin{tabular}{c|c|cccc cccc}
\hline
\multirow{2}{*}{Cluster} & \multirow{2}{*}{Efficiency Score} & \multirow{2}{*}{LSTM} & \multirow{2}{*}{KNN} & \multirow{2}{*}{1NN-DTW} & \multirow{2}{*}{HMM}
\\ & & & & & \\
\hline
\multirow{2}{*}{Top10Pr} 
& Eff. Gains (\%) & 2.61 & 2.13 & 3.20 & 6.05 \\
& IV Count (\#) & 134 & 114 & 127 & 139 \\
\hline
\multirow{2}{*}{Top25Pr} 
& Eff. Gains (\%) & 2.38 & 1.58 & 3.23 & 1.30 \\
& IV Count (\#) & 129 & 107 & 128 & 107 \\
\hline
\multirow{2}{*}{Top50Pr} 
& Eff. Gains (\%) & 0.97 & 0.98 & 2.58 & 7.34 \\
& IV Count (\#) & 100 & 106 & 117 & 140 \\
\hline
\multirow{2}{*}{Top75Pr} 
& Eff. Gains (\%) & -0.84 & 0.50 & 2.28 & 9.31 \\
& IV Count (\#) & 60 & 93 & 119 & 141 \\
\hline
\multirow{2}{*}{Average} 
& Eff. Gains (\%) & 1.28 & 1.30 & 2.82 & 6.00 \\
& IV Count (\#) & 105.75 & 105.00 & 122.75 & 131.75 \\
\hline
\end{tabular}
\end{table}

%%%%%%%%%%%%%%%
%%%%%%%%%%%%%%%
\begin{table}[h]
\centering
\caption{Average and deviation of gains (\%) of Eff-Scores in three weather states, when models are trained with the four data clusters}
\label{table_Stats_Effall_gains}
\begin{tabular}{|c|c c|c c|c c|c c|}
\hline
\multicolumn{1}{|c|}{Model} & \multicolumn{2}{|c|}{LSTM} & \multicolumn{2}{|c|}{KNN} & \multicolumn{2}{|c|}{1NN-DTW} & \multicolumn{2}{|c|}{HMM} \\
\hline
Weather & Avg & Std & Avg & Std & Avg & Std & Avg & Std \\
\hline
Calm & 0.17 & 4.72 & 1.17 & 3.29 & 2.21 & 4.30 & 3.96 & 6.19 \\
Moderate & 1.53 & 4.12 & 1.26 & 4.19 & 3.4 & 4.47 & 5.33 & 6.56 \\
Rough & 1.94 & 3.48 & 1.42 & 3.65 & 2.84 & 4.90 & 8.17 & 9.08 \\
\hline
Average & 1.21 & 4.11 & 1.28 & 3.71 & 2.82 & 4.56 & 5.82 & 7.28 \\
\end{tabular}
\end{table}

%%%%%%%%%%%%%%%%

As shown in Table~\ref{Table_Eff_gain_couts}, the HMM model achieves the highest average efficiency gain of 6.00\%, followed by the 1NN-DTW model (2.82\%), the KNN model (1.30\%), and the LSTM model (1.28\%). In terms of the number improved voyages out of 162 voyages in test dataset,  the HMM model also improves the energy efficiency of the most average number of improved voyages (131.75 out of 162 voyages).

The HMM model achieves its best performance when trained on the Top75Pr cluster, which includes voyages with lower Eff-Scores and frequently encountered adverse weather conditions. Such performance underscores the HMM model's capability to learn the hidden patterns between the vessel speed and weather states, ultimately facilitates for developing more efficient speed profiles.
%%%%%%%%%%%%%%

Table~\ref{table_Stats_Effall_gains} indicates that the HMM model yields the highest average efficiency gain in all three weather states, namely calm, moderate, and rough. The HMM model also has highest deviation of gains in all three weather states, indicating its adaptability to produce optimized speed profiles for more efficient voyages.

%%%%%%%%%%%%%%%%%%%%$  SOG and Eff-score Plots   %%%%%%%%%%%%

\begin{figure}[!ht]
\centering
\begin{subfigure}{0.48\linewidth}
    \includegraphics[width=\linewidth]{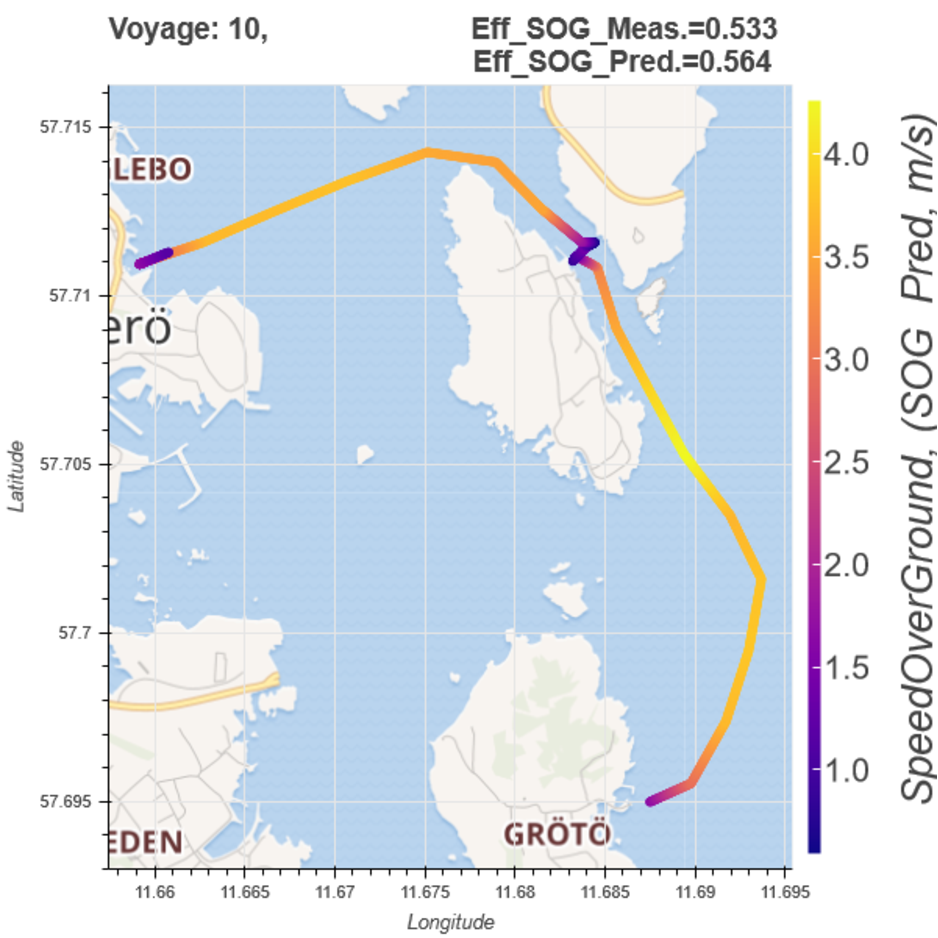}
    \caption{LSTM-based model}
    \label{voyage_LSTM}
\end{subfigure}
\hfill
\begin{subfigure}{0.48\linewidth}
    \includegraphics[width=\linewidth]{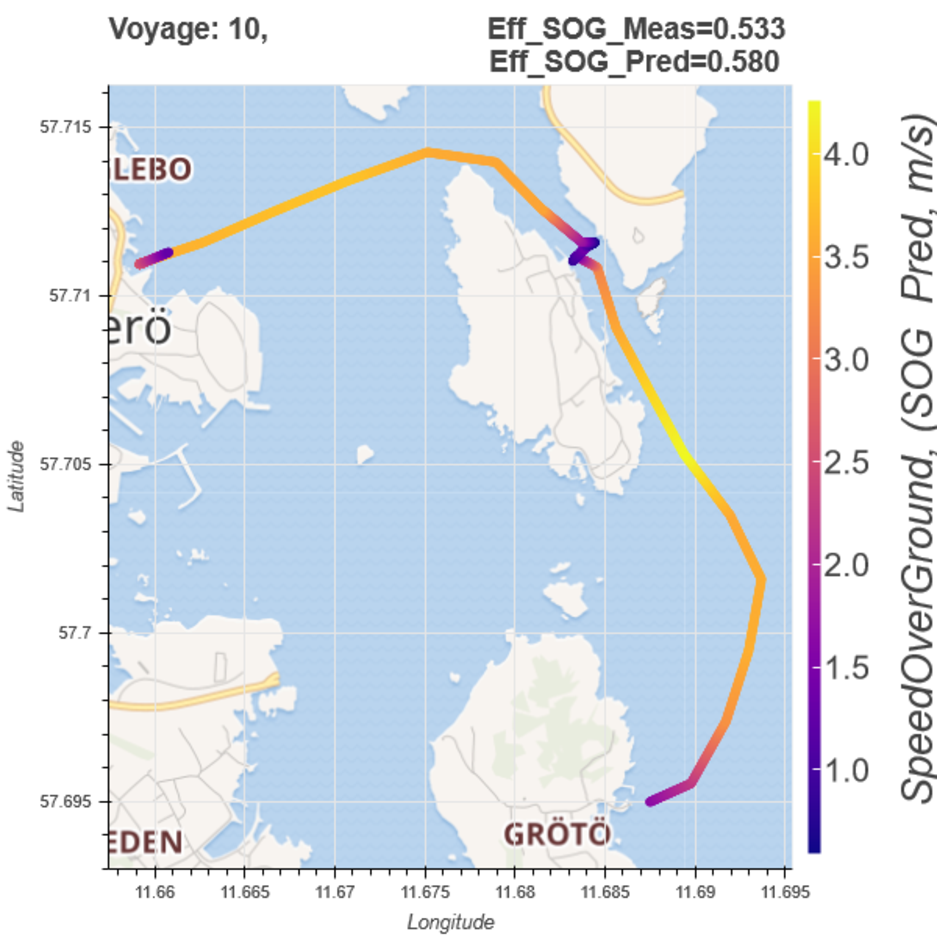}
    \caption{KNN-based model}
    \label{voyage_kNN}
\end{subfigure}
\hfill
\begin{subfigure}{0.48\linewidth}
    \includegraphics[width=\linewidth]{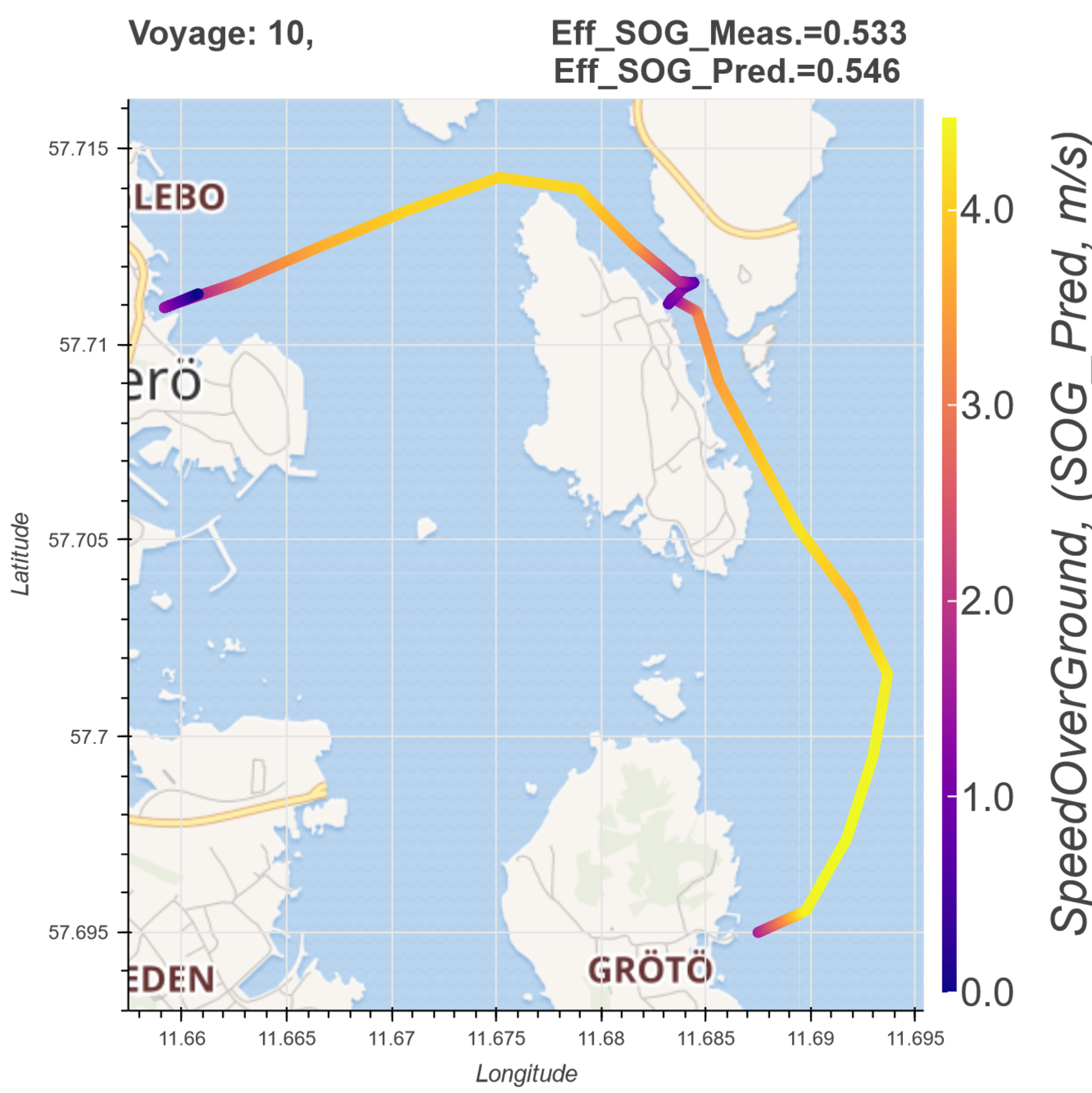}
    \caption{1NN-DTW-based model}
    \label{voyage_DTW}
\end{subfigure}
\hfill
\begin{subfigure}{0.48\linewidth}
    \includegraphics[width=\linewidth]{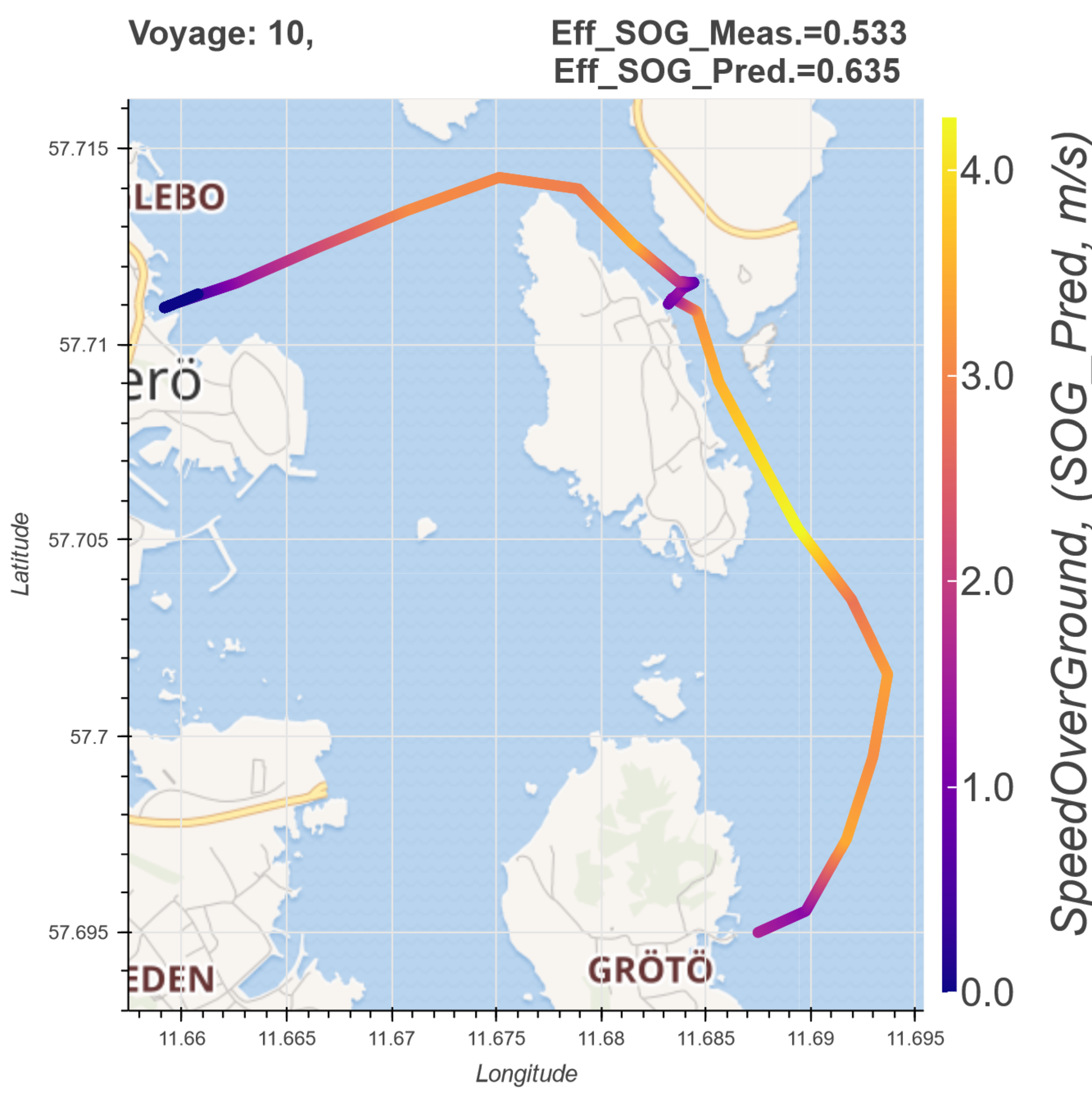}
    \caption{HMM-based model}
    \label{voyage_HMM}
\end{subfigure}
\caption{Predicted speed profile for a test voyage. From four time-series based models incorporate weather data as inputs and are trained by Top10Pr cluster.}
\label{Plots_sog_eff_seq}
\end{figure}
%%%%%%%%%%%%%%%

\begin{figure}[!ht]
\centering
\begin{subfigure}{0.49\linewidth}
    \includegraphics[width=\linewidth]{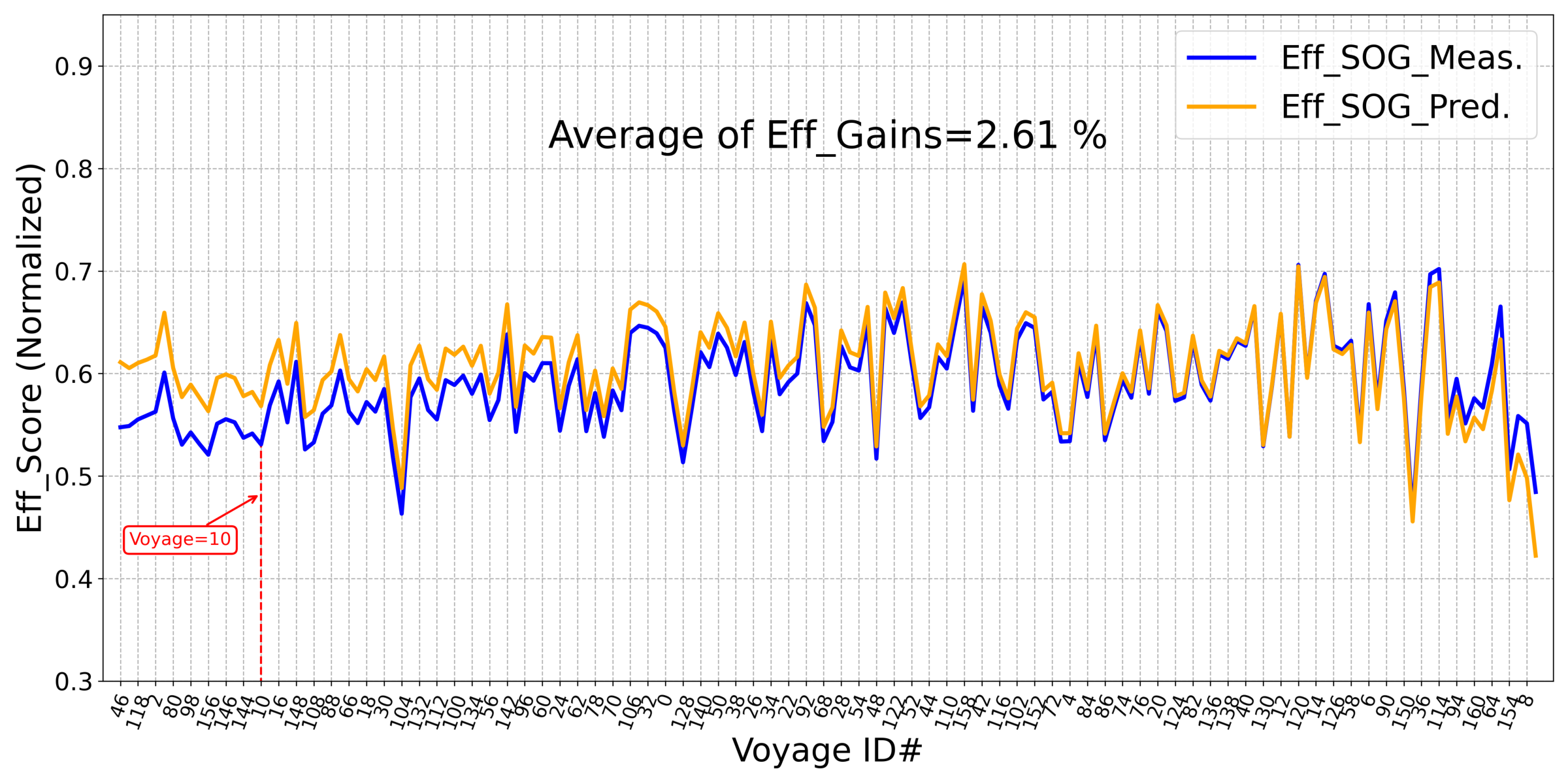}
    
    \caption{LSTM-based model}
    \label{Sort_Effs_graph_LSTM}
\end{subfigure}
\hfill
\begin{subfigure}{0.49\linewidth}
    \includegraphics[width=\linewidth]{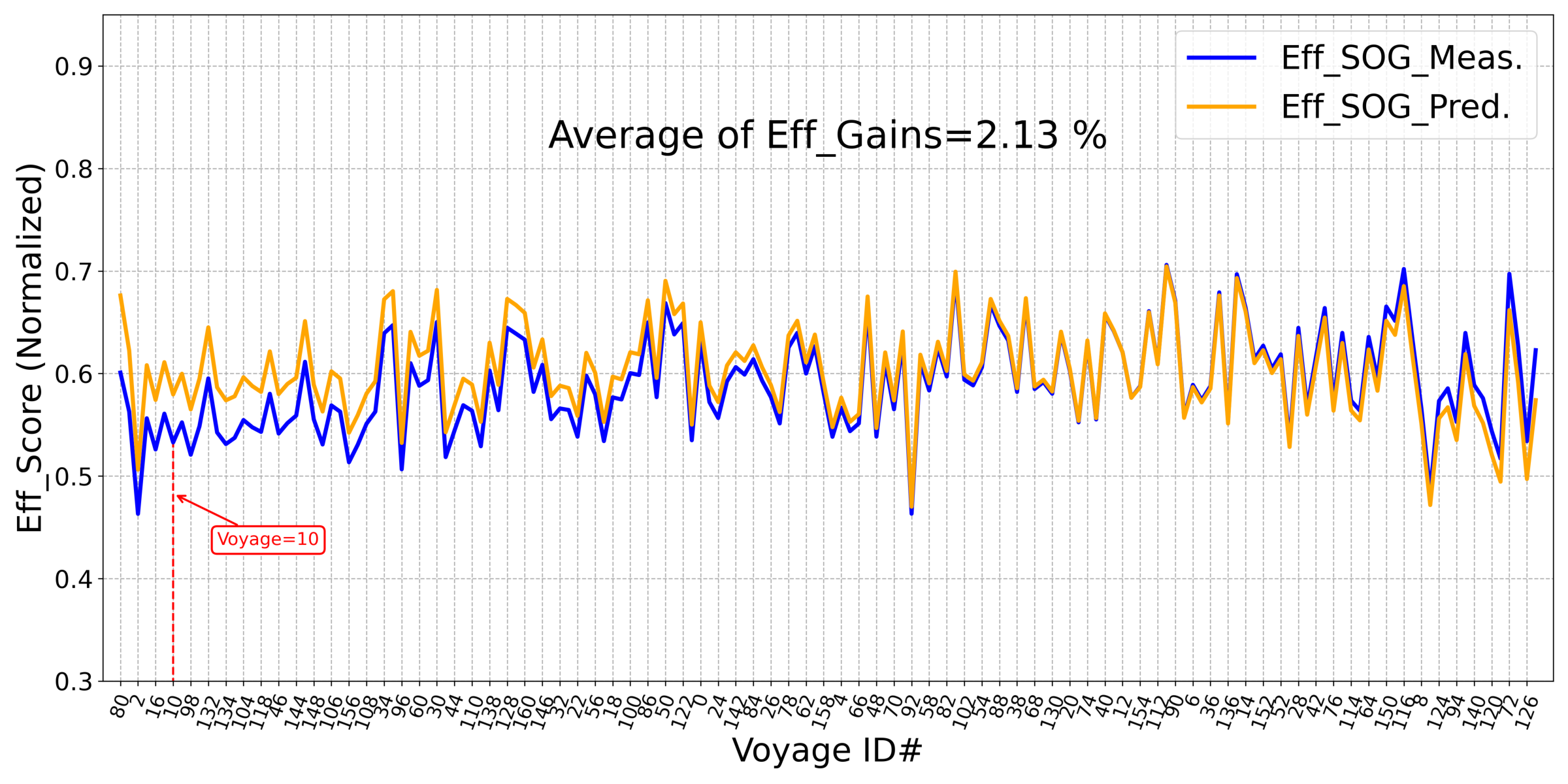}
    \caption{KNN-based model}
    \label{Sort_Effs_graph_kNN}
\end{subfigure}
\hfill
\begin{subfigure}{0.49\linewidth}
    \includegraphics[width=\linewidth]{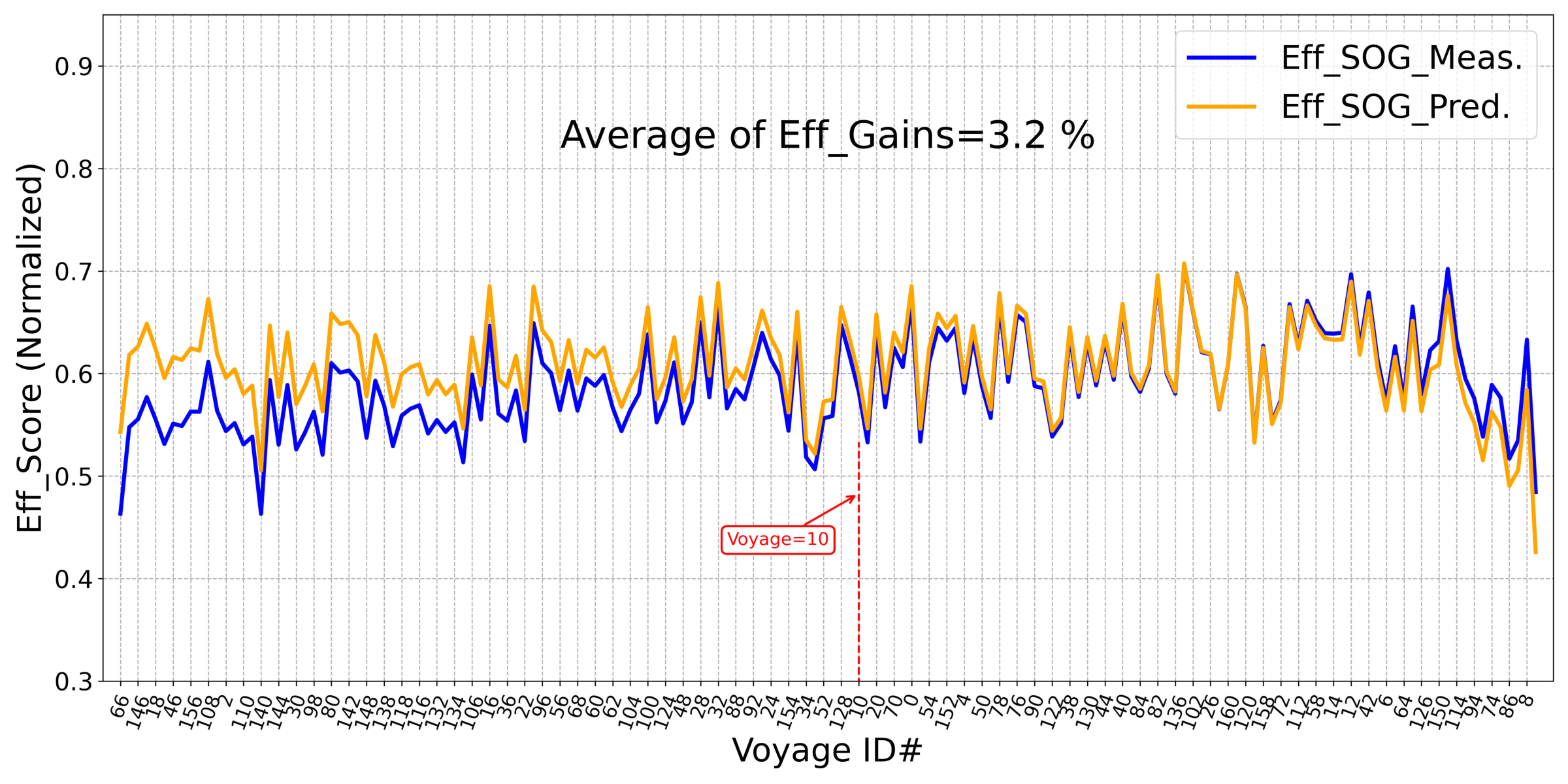}
    \caption{1NN-DTW-based model}
    \label{Sort_Effs_graph_DTW}
\end{subfigure}
\hfill
\begin{subfigure}{0.49\linewidth}
    \includegraphics[width=\linewidth]{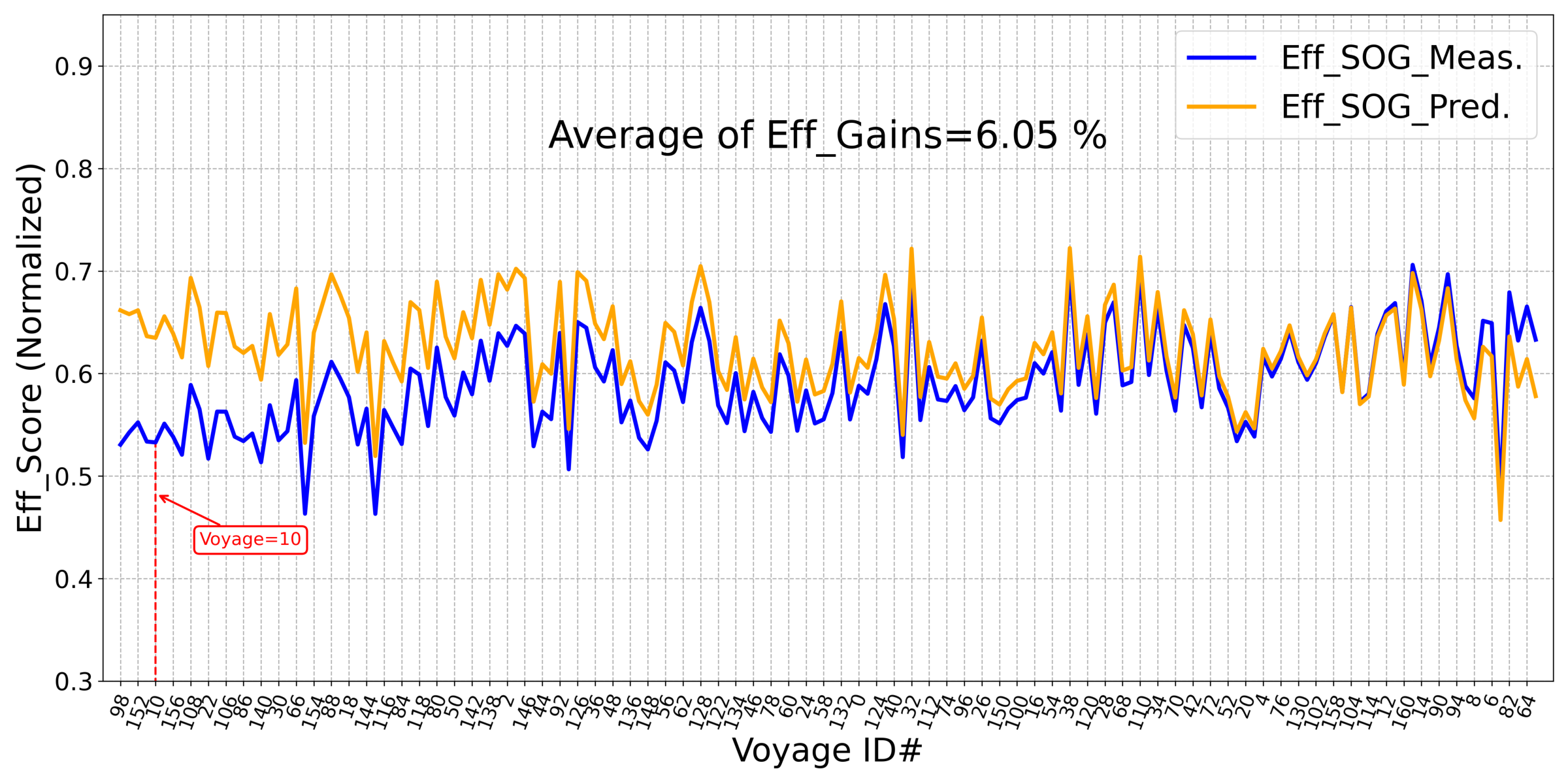}
    \caption{HMM-based model}
    \label{Sort_Effs_graph_HMM}
\end{subfigure}
\caption{Sorted Eff-Scores based on their gain, for 162 test voyages. From four time-series based models incorporate weather data as inputs and are trained by Top10Pr cluster,  
$Eff._{\text{Gain}}$ as in Eq.~\ref{eq:eff_gain}.}
\label{Plots_sort_effs}
\end{figure}
%%%%%%%%%%%%%%%%%%%%%%%%%%
We also present the results of our analysis through two key visualizations. The first plot illustrates the predicted speed profiles for a test voyage, which are generated by four time-series based models that incorporate weather data as inputs. These models were trained using data from the Top10Pr cluster. The second plot showcases the sorted Eff-Scores for 162 test voyages, based on their efficiency gains. These models, similar to the first plot, also integrate weather data as inputs and were trained by the Top10Pr cluster. The efficiency gain is quantified using the formula as defined in Eq.~\ref{eq:eff_gain}.
%%%%%%%%%%%%%%%%%%%%%%%%%%

In summary, the HMM-based model is the most effective model for improving energy efficiency for a vessel voyage in short sea. The HMM model is able to learn the complex relationships between the input features (e.g., speed and weather) and the output feature (Eff-Score), even in different weather conditions.

%%%%%%%%%%%%%%%%%%%%%%%%%%%%%%%%%%%%%%%%%%%%%%%%%%%%%%%%%%%%%%%%%
\section{Conclusion}\label{Sec_conclusion}
We have introduced a data-driven framework designed to optimize vessel speed profiles and enhance the energy efficiency of voyages in short-sea shipping (SSS). The framework integrates a data-driven approach, incorporating explainable artificial intelligence (XAI) for modeling vessel energy, along with a time-series analysis model for optimizing speed profiles.

Four time-series analysis models, namely LSTM, KNN, 1NN-DTW, and HMM, were employed to analyze and evaluate the framework's effectiveness using real-world data. Our findings underscore the framework's capability to significantly improve vessel energy efficiency, especially when faced with limited options as in SSS.

The results highlight the superiority of the framework with the HMM model over LSTM, KNN and 1NN-DTW. Particularly, the HMM model demonstrates enhanced performance when trained with data including voyages with diverse efficiency levels and frequent exposure to different weather conditions.

In summary, the proposed framework, especially when incorporating the HMM model, consistently yields efficiency gains, showcasing its adaptability in generating optimized speed profiles tailored to varying environmental conditions. This establishes the HMM model as a robust choice among time-series analysis models, effectively improving energy efficiency in maritime voyages facing challenges and constrained options, such scenarios are typical in SSS.

\section*{Acknowledgment}
This research project is funded by Sweden’s innovation agency (Vinnova).\\
% We would like to thank CetaSol AB for their support and for providing the resources necessary to conduct this research.\\
The authors wish to thank the diverse group at the Center for Applied Intelligent Systems Research (CAISR), Halmstad University, for helpful discussions.

% {\appendix[Proof of the Zonklar Equations]
% Use $\backslash${\tt{appendix}} if you have a single appendix:
% Do not use $\backslash${\tt{section}} anymore after $\backslash${\tt{appendix}}, only $\backslash${\tt{section*}}.
% If you have multiple appendixes use $\backslash${\tt{appendices}} then use $\backslash${\tt{section}} to start each appendix.
% You must declare a $\backslash${\tt{section}} before using any $\backslash${\tt{subsection}} or using $\backslash${\tt{label}} ($\backslash${\tt{appendices}} by itself starts a section numbered zero.)}
 
{\appendix[Supplementary Materials]
The source codes that are implemented on Python 3.9.7 to produce the results are available at: \url{https://github.com/MohamedAbuella/TSA4EESSS}
The dataset is private and cannot be shared due to the crucial commercial interests of the startup company operating the iHelm system.}

%% If this paper submitted to a double-blind review, then the github should be anonymous, but we have already mentioned the previous study of using XAI for voyage efficiency modeling!
%% \url{https://anonymous .4open.science/r/TSA4EESSS....}

% \section{References Section}

\bibliographystyle{IEEEtran}
\bibliography{ihelm_bib}

\begin{IEEEbiography}[{\includegraphics[width=1in,height=1.25in,clip,keepaspectratio]{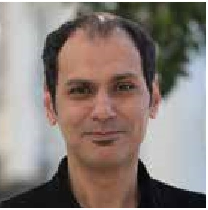}}]
{Mohamed Abuella} received his M.S. and PhD degrees in Electrical and Computer Engineering from Southern Illinois University at Carbondale and University of North Carolina at Charlotte, in 2012 and 2018 respectively. He is a postdoctoral researcher at Halmstad University since 2022.
His research interests include energy analytics and AI for sustainability.
\end{IEEEbiography}

\begin{IEEEbiography}[{\includegraphics[width=1in,height=1.25in,clip,keepaspectratio]{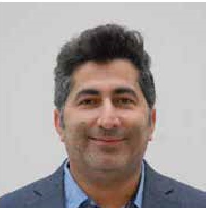}}]
{Hadi Fanaee-T} is an Associate Professor at Halmstad University, Sweden. He completed his PhD (with distinction) in Computer Science, under supervision of Professor Joao Gama at the Faculty of Science of University of Porto, Portugal in November 2015. He was the finalist in ERCIM Cor Baayen Young Researcher Award 2017. Prior to this position he was a postdoctoral fellow at the department of biostatistics, University of Oslo, Norway, and also worked as a postdoctoral researcher in European FP7 Project "MAESTRA" at INESC TEC research institute, Portugal. His main research interests are interdisciplinary applications of tensor decompositions, data fusion, anomaly/event detection and spatiotemporal data mining. He is the first-author of several journal and conference papers. He has served as a PC member to over 20 prestigious conferences (e.g. IJCAI, AAAI, ECML-PKDD, ISMIS, ACM SAC, IEEE DSAA, etc.) and also reviewer to several high-impact journals (e.g. TDKE, TKDD, DAMI, ML, KAIS, KBS, CSUR.
\end{IEEEbiography}

\begin{IEEEbiography}[{\includegraphics[width=1in,height=1.25in,clip,keepaspectratio]{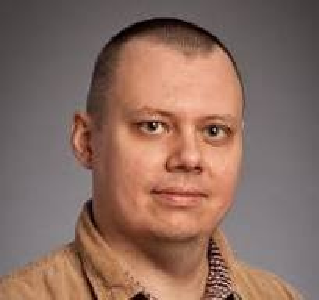}}]
{Slawomir Nowaczyk} is a Professor in Machine Learning at the Center for Applied Intelligent Systems Research, Halmstad University, Sweden. He received his MSc degree from Poznan University of Technology in 2002 and his PhD from the Lund University of Technology in 2008. During the last decades, his research has focused on machine learning, knowledge representation, and self-organising systems. The majority of his work concerns industrial data streams, often with predictive maintenance as the goal. Given that accurate and relevant labels are usually impossible to obtain in such settings, Slawomir’s contributions primarily take advantage of proxy labels, such as transfer learning and multi-task learning, or concern semi-supervised and unsupervised modelling. He is a board member of the Swedish AI Society and a research leader for the School of Information Technology at Halmstad University. Slawomir has led multiple research projects on applying Artificial Intelligence and Machine Learning in different domains, such as transport and automotive, energy, smart cities, and healthcare. In most cases, this research was done in collaboration with industry and public administration organisations – inspired by practical challenges and leading to tangible results and deployed solutions.
\end{IEEEbiography}

\begin{IEEEbiography}[{\includegraphics[width=1in,height=1.25in,clip,keepaspectratio]{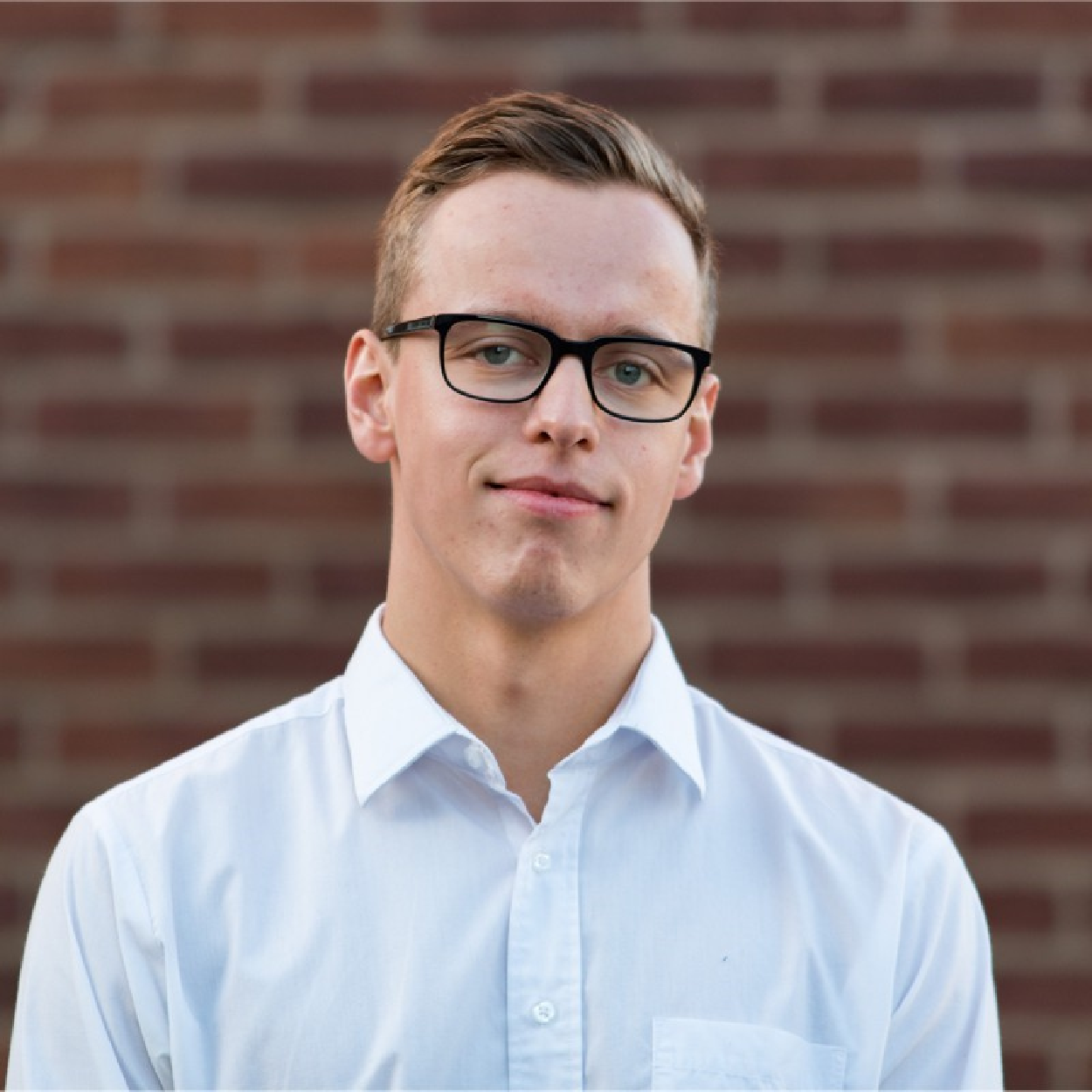}}]
{Simon Johansson} is a MSc graduate of Chalmers University of Technology's program in Engineering Mathematics and Computational Science in 2020, currently works in Cetasol, a marine company specialising in CO2 reduction and energy optimisation for vessels. His practical application of computational methods and dedication to environmental sustainability align with his role, contributing to global efforts to mitigate climate change. Simon's commitment to advancing eco-friendly solutions in the marine industry reflects a seamless integration of academic excellence and real-world impact.
\end{IEEEbiography}

\begin{IEEEbiography}[{\includegraphics[width=1in,height=1.25in,clip,keepaspectratio]{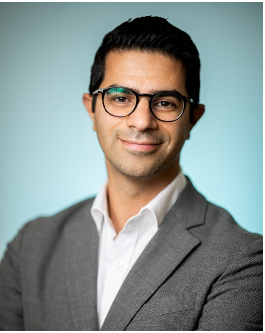}}]
{Ethan Faghani} is the CEO and founder of Cetasol. Before Cetasol, Ethan was Chief Engineer of Automation and AI at Volvo Penta. Ethan has experience working with cutting-edge technologies in other transportation segments in both big enterprises and his own founded startup. Ethan obtained his Ph.D. in mechatronics from UBC and Innovation and Entrepreneurship from Stanford Business School.
\end{IEEEbiography}

% \newpage

 

% \end{IEEEbiography}



\vfill

\end{document}